\documentclass[AMA,STIX2COL]{MRM}

\articletype{Pre-Print -- submitted to NMR in Biomedicine}%

\received{\today}
\revised{}
\accepted{}
\usepackage{tikz}
\usepackage{pgfplots}

\usetikzlibrary{spy,backgrounds,arrows,decorations.pathmorphing,backgrounds,positioning,fit,matrix,calc,shapes.multipart,shapes.misc,pgfplots.statistics, pgfplots.colorbrewer}

\pgfplotsset{compat=1.18,
	boxplot/estimator=R1,
	colormap={parula}{
			rgb255=(53,42,135)
			rgb255=(15,92,221)
			rgb255=(18,125,216)
			rgb255=(7,156,207)
			rgb255=(21,177,180)
			rgb255=(89,189,140)
			rgb255=(165,190,107)
			rgb255=(225,185,82)
			rgb255=(252,206,46)
			rgb255=(249,251,14)},
	colormap={mybluered}{
			rgb255(0cm)=(0,0,180)
			rgb255(1cm)=(0,180,180)
			rgb255(2cm)=(70,180,0)
			rgb255(3cm)=(180,180,0)
			rgb255(4cm)=(255,0,0)
			rgb255(5cm)=(128,0,0)},
	colormap={mybluewhitered_m10_2}{
			rgb255(0cm)=(0,0,255)
			rgb255(10cm)=(255,255,255)
			rgb255(12cm)=(255,0,0)},
	colormap={mybluewhitered_m1_2}{
			rgb255(0cm)=(0,0,255)
			rgb255(1cm)=(255,255,255)
			rgb255(3cm)=(255,0,0)},
	colormap={mybluewhitered_0_1}{
			rgb255(0cm)=(255,255,255)
			rgb255(1cm)=(255,0,0)},
	colormap={gist_earth}{
			rgb255=(0.0,0.0,0.0)
			rgb255=(1.7085000000000001,0.0,62.424)
			rgb255=(3.3914999999999997,0.0,89.8365)
			rgb255=(5.1000000000000005,1.4789999999999999,113.985)
			rgb255=(6.8085,7.343999999999999,116.4075)
			rgb255=(8.4915,13.209,116.8665)
			rgb255=(10.200000000000001,19.074,117.3255)
			rgb255=(11.9085,24.939,117.7845)
			rgb255=(13.5915,30.804000000000002,118.2435)
			rgb255=(15.299999999999999,36.669000000000004,118.7025)
			rgb255=(16.983,42.534,119.1615)
			rgb255=(18.6915,48.3735,119.6205)
			rgb255=(20.400000000000002,53.677499999999995,120.10499999999999)
			rgb255=(22.083,58.981500000000004,120.564)
			rgb255=(23.7915,64.2855,121.02300000000001)
			rgb255=(25.5,69.58949999999999,121.482)
			rgb255=(27.183,74.8935,121.941)
			rgb255=(28.8915,79.917,122.39999999999999)
			rgb255=(30.599999999999998,84.6855,122.859)
			rgb255=(32.282999999999994,89.454,123.3435)
			rgb255=(33.9915,94.2225,123.8025)
			rgb255=(35.7,98.838,124.2615)
			rgb255=(37.383,102.86699999999999,124.7205)
			rgb255=(39.091499999999996,106.896,125.1795)
			rgb255=(40.774499999999996,110.925,125.63850000000001)
			rgb255=(42.483,114.954,126.0975)
			rgb255=(44.191500000000005,118.983,126.5565)
			rgb255=(45.8745,123.012,127.041)
			rgb255=(47.583,127.041,127.5)
			rgb255=(48.909,129.13199999999998,125.23049999999999)
			rgb255=(50.1075,130.5855,122.1195)
			rgb255=(51.306,132.0645,118.983)
			rgb255=(52.5045,133.518,115.872)
			rgb255=(53.7285,134.99699999999999,112.7355)
			rgb255=(54.927,136.476,109.6245)
			rgb255=(56.125499999999995,137.92950000000002,106.51350000000001)
			rgb255=(57.324,139.4085,103.377)
			rgb255=(58.5225,140.862,100.266)
			rgb255=(59.721,142.341,97.12950000000001)
			rgb255=(60.945,143.7945,94.0185)
			rgb255=(62.1435,145.27349999999998,90.882)
			rgb255=(63.342000000000006,146.7525,87.771)
			rgb255=(64.5405,148.20600000000002,84.66000000000001)
			rgb255=(65.73899999999999,149.685,81.5235)
			rgb255=(66.9375,151.1385,78.4125)
			rgb255=(68.136,152.6175,75.27600000000001)
			rgb255=(69.8955,154.071,72.16499999999999)
			rgb255=(75.5565,155.54999999999998,70.5585)
			rgb255=(81.2175,157.029,72.063)
			rgb255=(86.904,158.48250000000002,73.542)
			rgb255=(92.565,159.9615,75.021)
			rgb255=(98.226,161.415,76.5)
			rgb255=(103.887,162.894,78.00450000000001)
			rgb255=(109.57350000000001,164.06699999999998,79.48349999999999)
			rgb255=(115.23450000000001,165.18900000000002,80.9625)
			rgb255=(120.74249999999999,166.311,82.11)
			rgb255=(124.95,167.4075,82.926)
			rgb255=(129.1575,168.5295,83.742)
			rgb255=(133.365,169.6515,84.5325)
			rgb255=(137.5725,170.77349999999998,85.3485)
			rgb255=(141.78,171.8955,86.16449999999999)
			rgb255=(145.9875,173.01749999999998,86.95500000000001)
			rgb255=(150.195,174.1395,87.771)
			rgb255=(154.4025,175.2615,88.5615)
			rgb255=(158.60999999999999,176.3835,89.3775)
			rgb255=(162.792,177.48,90.1935)
			rgb255=(166.9995,178.602,90.984)
			rgb255=(171.207,179.724,91.8)
			rgb255=(175.41449999999998,180.846,92.59049999999999)
			rgb255=(179.622,181.968,93.40650000000001)
			rgb255=(183.192,182.42700000000002,94.2225)
			rgb255=(184.263,180.2085,95.01299999999999)
			rgb255=(185.334,178.0155,95.82900000000001)
			rgb255=(186.405,175.8225,96.6195)
			rgb255=(187.476,173.60399999999998,97.4355)
			rgb255=(188.5215,171.411,98.2515)
			rgb255=(189.5925,169.218,99.042)
			rgb255=(190.6635,166.9995,99.858)
			rgb255=(191.7345,164.8065,100.6485)
			rgb255=(193.1115,163.2255,104.5755)
			rgb255=(195.9675,164.3985,110.823)
			rgb255=(198.8235,165.75,117.0705)
			rgb255=(201.67950000000002,167.94299999999998,123.318)
			rgb255=(204.5355,170.1615,129.5655)
			rgb255=(207.366,172.38000000000002,135.813)
			rgb255=(210.222,174.573,142.06050000000002)
			rgb255=(213.078,176.7915,148.308)
			rgb255=(215.934,179.469,154.5555)
			rgb255=(218.79,183.141,160.80300000000003)
			rgb255=(221.646,186.66,167.0505)
			rgb255=(224.4765,190.4595,174.1395)
			rgb255=(227.33249999999998,195.45749999999998,181.815)
			rgb255=(230.18849999999998,200.43,189.516)
			rgb255=(233.0445,205.428,197.1915)
			rgb255=(235.90050000000002,210.42600000000002,204.89249999999998)
			rgb255=(238.75650000000002,216.1635,212.568)
			rgb255=(241.6125,222.0285,220.2435)
			rgb255=(244.443,228.7605,227.9445)
			rgb255=(247.299,236.2065,235.62)
			rgb255=(250.155,243.6525,243.32100000000003)
			rgb255=(253.011,250.9965,250.9965)
		},
	colormap={berlin}{
			rgb255=(158.37591,175.99641,254.87428500000001)
			rgb255=(156.100035,175.75313999999997,253.82037)
			rgb255=(153.81651,175.50375,252.765945)
			rgb255=(151.521,175.250535,251.70846)
			rgb255=(149.21707500000002,174.99324,250.64511)
			rgb255=(146.90244,174.73161,249.57691499999999)
			rgb255=(144.57505500000002,174.46233,248.50361999999998)
			rgb255=(142.236705,174.18999,247.42344)
			rgb255=(139.89045000000002,173.90796,246.33408)
			rgb255=(137.52838500000001,173.619045,245.23452)
			rgb255=(135.15867,173.32120500000002,244.12425000000002)
			rgb255=(132.775185,173.01342,243.00020999999998)
			rgb255=(130.38022500000002,172.69365,241.85883)
			rgb255=(127.97506499999999,172.35654,240.70036499999998)
			rgb255=(125.55384,172.00413,239.520735)
			rgb255=(123.12216,171.63412499999998,238.31637)
			rgb255=(120.675945,171.24015,237.086505)
			rgb255=(118.22055,170.82399,235.82553000000001)
			rgb255=(115.752405,170.37876,234.532425)
			rgb255=(113.274315,169.90012499999997,233.20209000000003)
			rgb255=(110.7822,169.38910499999997,231.82968000000002)
			rgb255=(108.284475,168.84060000000002,230.41443)
			rgb255=(105.77859000000001,168.246705,228.950475)
			rgb255=(103.268625,167.608695,227.43577499999998)
			rgb255=(100.759935,166.917645,225.86625)
			rgb255=(98.25048000000001,166.17686999999998,224.23884)
			rgb255=(95.750715,165.37668,222.553035)
			rgb255=(93.26421,164.51682,220.805265)
			rgb255=(90.795045,163.59576,218.994255)
			rgb255=(88.345515,162.61146000000002,217.120005)
			rgb255=(85.93041,161.56136999999998,215.18302500000001)
			rgb255=(83.54871,160.443195,213.184335)
			rgb255=(81.214185,159.260505,211.12393500000002)
			rgb255=(78.932445,158.012535,209.00514)
			rgb255=(76.69992,156.696735,206.83254000000002)
			rgb255=(74.538795,155.324325,204.606645)
			rgb255=(72.44499,153.888675,202.33485)
			rgb255=(70.43227499999999,152.39667,200.01843)
			rgb255=(68.49172499999999,150.8529,197.661465)
			rgb255=(66.63354,149.26042500000003,195.2739)
			rgb255=(64.86384,147.62153999999998,192.85548)
			rgb255=(63.177015,145.94364,190.413345)
			rgb255=(61.57332,144.22698,187.95183)
			rgb255=(60.056325,142.48074,185.474505)
			rgb255=(58.60971,140.70951,182.988)
			rgb255=(57.248265,138.91125,180.490275)
			rgb255=(55.968675000000005,137.09514,177.98949)
			rgb255=(54.74697,135.26041500000002,175.48845)
			rgb255=(53.59386,133.414215,172.986135)
			rgb255=(52.501695,131.55373500000002,170.48739)
			rgb255=(51.451605,129.69249,167.990685)
			rgb255=(50.45889,127.82078999999999,165.50265)
			rgb255=(49.51386,125.94526499999999,163.01818500000002)
			rgb255=(48.59178,124.067955,160.54086)
			rgb255=(47.71356,122.19115500000001,158.07398999999998)
			rgb255=(46.85676,120.31563000000001,155.611455)
			rgb255=(46.027499999999996,118.44087,153.15861)
			rgb255=(45.228075,116.564835,150.714435)
			rgb255=(44.43732,114.69594000000001,148.278165)
			rgb255=(43.662119999999994,112.83086999999999,145.85133)
			rgb255=(42.90171,110.96886,143.43393)
			rgb255=(42.14946,109.11041999999999,141.020355)
			rgb255=(41.421945,107.25504,138.618765)
			rgb255=(40.683975,105.40935,136.22355)
			rgb255=(39.968444999999996,103.56748499999999,133.83828)
			rgb255=(39.245774999999995,101.72766,131.45862)
			rgb255=(38.53611,99.89803500000001,129.08865)
			rgb255=(37.828230000000005,98.070705,126.72786)
			rgb255=(37.138455,96.25281000000001,124.376505)
			rgb255=(36.434145,94.43746499999999,122.02872)
			rgb255=(35.73519,92.630535,119.694195)
			rgb255=(35.05383,90.83202,117.360945)
			rgb255=(34.368135,89.03630999999999,115.042485)
			rgb255=(33.680145,87.24825,112.731675)
			rgb255=(32.997254999999996,85.469115,110.42571)
			rgb255=(32.317425,83.68972500000001,108.12918)
			rgb255=(31.642950000000003,81.921045,105.84310500000001)
			rgb255=(30.97128,80.158485,103.56672)
			rgb255=(30.319245000000002,78.4023,101.29467)
			rgb255=(29.66058,76.65504,99.03588)
			rgb255=(29.001405,74.91415500000001,96.78423)
			rgb255=(28.352684999999997,73.1799,94.54074)
			rgb255=(27.696315000000002,71.455335,92.31102)
			rgb255=(27.070545,69.74173499999999,90.0864)
			rgb255=(26.43585,68.02788000000001,87.87147)
			rgb255=(25.801665,66.32754,85.66775999999999)
			rgb255=(25.18788,64.634085,83.47221)
			rgb255=(24.568485,62.94675,81.289665)
			rgb255=(23.985045,61.26732,79.11808500000001)
			rgb255=(23.405939999999998,59.600384999999996,76.94829)
			rgb255=(22.82403,57.947475,74.79609)
			rgb255=(22.271955000000002,56.304,72.65307)
			rgb255=(21.711209999999998,54.6618,70.52687999999999)
			rgb255=(21.182595000000003,53.035155,68.403495)
			rgb255=(20.67999,51.415905,66.29796)
			rgb255=(20.178150000000002,49.817055,64.204155)
			rgb255=(19.707929999999998,48.22968,62.115195)
			rgb255=(19.270605,46.650465,60.05301)
			rgb255=(18.868215,45.092925,57.99567)
			rgb255=(18.46455,43.550174999999996,55.955414999999995)
			rgb255=(18.116474999999998,42.022725,53.9325)
			rgb255=(17.790585,40.519754999999996,51.92514)
			rgb255=(17.49759,39.025200000000005,49.933589999999995)
			rgb255=(17.2278,37.566345,47.97162)
			rgb255=(16.999575,36.126104999999995,46.01526)
			rgb255=(16.810364999999997,34.69938,44.093835)
			rgb255=(16.661444999999997,33.327225,42.19179)
			rgb255=(16.552305,31.963994999999997,40.333095)
			rgb255=(16.48218,30.63366,38.49123)
			rgb255=(16.451835000000003,29.342850000000002,36.691694999999996)
			rgb255=(16.46127,28.10661,34.924034999999996)
			rgb255=(16.510995,26.883885000000003,33.208650000000006)
			rgb255=(16.6005,25.716495,31.510095)
			rgb255=(16.672665,24.599595,29.878349999999998)
			rgb255=(16.721369999999997,23.54619,28.30704)
			rgb255=(16.80246,22.491255,26.770410000000002)
			rgb255=(16.92894,21.45417,25.31844)
			rgb255=(17.11254,20.413005,23.926395)
			rgb255=(17.389215,19.405245,22.559849999999997)
			rgb255=(17.7786,18.432165,21.171375)
			rgb255=(18.267944999999997,17.506770000000003,19.77372)
			rgb255=(18.86439,16.58979,18.38805)
			rgb255=(19.53198,15.722534999999999,16.996005)
			rgb255=(20.307434999999998,14.93025,15.588915)
			rgb255=(21.155565,14.19483,14.214975)
			rgb255=(22.066935,13.514235000000001,12.83568)
			rgb255=(23.030325,12.928245,11.485199999999999)
			rgb255=(24.0363,12.432015,10.142115)
			rgb255=(25.071345,11.995455,8.844165)
			rgb255=(26.12679,11.63412,7.66887)
			rgb255=(27.216659999999997,11.399775,6.63306)
			rgb255=(28.30143,11.21286,5.706645)
			rgb255=(29.387475,11.11698,4.88325)
			rgb255=(30.484485,11.109585000000001,4.156245)
			rgb255=(31.57206,11.184555,3.5182350000000002)
			rgb255=(32.666775,11.337045,2.9549399999999997)
			rgb255=(33.740325,11.533394999999999,2.430405)
			rgb255=(34.795004999999996,11.77182,2.013225)
			rgb255=(35.861925,12.08037,1.65801)
			rgb255=(36.945420000000006,12.40167,1.358385)
			rgb255=(38.05365,12.708179999999999,1.10823)
			rgb255=(39.189674999999994,13.004235,0.901935)
			rgb255=(40.368795,13.29315,0.73491)
			rgb255=(41.56857,13.570590000000001,0.602565)
			rgb255=(42.791805,13.831199999999999,0.5005649999999999)
			rgb255=(44.04768,14.068859999999999,0.425595)
			rgb255=(45.339254999999994,14.28459,0.374595)
			rgb255=(46.630065,14.4891,0.3417)
			rgb255=(47.95479,14.68137,0.32181000000000004)
			rgb255=(49.274415,14.921069999999999,0.31263)
			rgb255=(50.608065,15.18525,0.312885)
			rgb255=(51.96339,15.427755,0.32130000000000003)
			rgb255=(53.31846,15.67893,0.33711)
			rgb255=(54.68985,15.991050000000001,0.36006)
			rgb255=(56.073735,16.274865000000002,0.389895)
			rgb255=(57.462975,16.581885,0.42712500000000003)
			rgb255=(58.86828,16.905735,0.472515)
			rgb255=(60.28761,17.249475,0.52734)
			rgb255=(61.71408,17.61846,0.5928749999999999)
			rgb255=(63.158655,17.966790000000003,0.67116)
			rgb255=(64.61445,18.35643,0.76449)
			rgb255=(66.08988000000001,18.7782,0.8759250000000001)
			rgb255=(67.574235,19.185435,1.0085250000000001)
			rgb255=(69.08817,19.626075,1.165605)
			rgb255=(70.61664,20.099610000000002,1.351755)
			rgb255=(72.169335,20.608845,1.5710549999999999)
			rgb255=(73.74498,21.134145,1.828605)
			rgb255=(75.34383,21.694125,2.128995)
			rgb255=(76.97379000000001,22.3023,2.48013)
			rgb255=(78.63868500000001,22.92756,2.9210249999999998)
			rgb255=(80.33112000000001,23.59515,3.3976200000000003)
			rgb255=(82.06053,24.32496,3.930315)
			rgb255=(83.82818999999999,25.09098,4.5339)
			rgb255=(85.63027500000001,25.902900000000002,5.214494999999999)
			rgb255=(87.47418,26.769135,5.9772)
			rgb255=(89.35531499999999,27.7032,6.826605)
			rgb255=(91.27648500000001,28.70382,7.76628)
			rgb255=(93.235395,29.74779,8.815605)
			rgb255=(95.23485000000001,30.847604999999998,9.974324999999999)
			rgb255=(97.273065,32.02953,11.141715000000001)
			rgb255=(99.343665,33.266535,12.360105)
			rgb255=(101.450475,34.54587,13.54968)
			rgb255=(103.5861,35.902725000000004,14.75124)
			rgb255=(105.74595000000001,37.29987,15.992325000000001)
			rgb255=(107.923395,38.754645000000004,17.259674999999998)
			rgb255=(110.118435,40.26603,18.62622)
			rgb255=(112.31322,41.82714,20.048099999999998)
			rgb255=(114.516675,43.418595,21.58422)
			rgb255=(116.71452,45.04983,23.171595)
			rgb255=(118.91007,46.719314999999995,24.820425)
			rgb255=(121.09949999999999,48.42144,26.53632)
			rgb255=(123.27210000000001,50.152635,28.314944999999998)
			rgb255=(125.43705,51.89658,30.138450000000002)
			rgb255=(127.58211000000001,53.660415,32.002755)
			rgb255=(129.71595,55.443375,33.910664999999995)
			rgb255=(131.829135,57.23016,35.858865)
			rgb255=(133.92523500000001,59.043465,37.828995)
			rgb255=(136.003995,60.84912,39.846555)
			rgb255=(138.0672,62.667525,41.879414999999995)
			rgb255=(140.117655,64.495365,43.927575)
			rgb255=(142.15281000000002,66.323205,46.002765000000004)
			rgb255=(144.17139,68.150025,48.1032)
			rgb255=(146.18436,69.987555,50.215619999999994)
			rgb255=(148.18356,71.826615,52.335435000000004)
			rgb255=(150.1746,73.66797,54.474375)
			rgb255=(152.16258000000002,75.50907,56.62377)
			rgb255=(154.14342,77.352975,58.784895000000006)
			rgb255=(156.11814,79.207335,60.958259999999996)
			rgb255=(158.09387999999998,81.056085,63.14259)
			rgb255=(160.06554,82.90866,65.328195)
			rgb255=(162.03669,84.772965,67.527825)
			rgb255=(164.009115,86.63497500000001,69.73995000000001)
			rgb255=(165.983835,88.50131999999999,71.955645)
			rgb255=(167.95855500000002,90.37072500000001,74.176185)
			rgb255=(169.936845,92.246505,76.40820000000001)
			rgb255=(171.91947,94.12585499999999,78.645825)
			rgb255=(173.903625,96.01209,80.890845)
			rgb255=(175.894665,97.89959999999999,83.14096500000001)
			rgb255=(177.88698,99.79526999999999,85.40689499999999)
			rgb255=(179.88567,101.69247,87.66798)
			rgb255=(181.88844,103.599105,89.94359999999999)
			rgb255=(183.89529,105.50803499999999,92.22381)
			rgb255=(185.90877,107.42104499999999,94.50759000000001)
			rgb255=(187.92684,109.34298,96.80208)
			rgb255=(189.9495,111.26899499999999,99.102945)
			rgb255=(191.97700500000002,113.200365,101.4084)
			rgb255=(194.011905,115.13556000000001,103.72048500000001)
			rgb255=(196.049355,117.07661999999999,106.03869)
			rgb255=(198.09521999999998,119.02278,108.36531000000001)
			rgb255=(200.144145,120.973785,110.69345999999999)
			rgb255=(202.198425,122.9304,113.033595)
			rgb255=(204.257295,124.889565,115.378575)
			rgb255=(206.32305,126.85893,117.72636000000001)
			rgb255=(208.39161000000001,128.82778499999998,120.08205)
			rgb255=(210.46629000000001,130.80531000000002,122.44335)
			rgb255=(212.546835,132.78665999999998,124.808475)
			rgb255=(214.63146,134.774385,127.184565)
			rgb255=(216.720675,136.765425,129.56448)
			rgb255=(218.81346,138.761055,131.94924)
			rgb255=(220.91262,140.763315,134.340375)
			rgb255=(223.01484,142.769145,136.73559)
			rgb255=(225.121395,144.779055,139.13896499999998)
			rgb255=(227.23305,146.79585,141.54591)
			rgb255=(229.348785,148.814175,143.96025)
			rgb255=(231.46758,150.84015,146.37969)
			rgb255=(233.587905,152.87046,148.80576)
			rgb255=(235.71384,154.905615,151.239225)
			rgb255=(237.84207,156.9423,153.67549499999998)
			rgb255=(239.97438,158.988165,156.118395)
			rgb255=(242.10898500000002,161.03556,158.56716)
			rgb255=(244.24716,163.088565,161.024085)
			rgb255=(246.38584500000002,165.14514,163.48381500000002)
			rgb255=(248.527845,167.20809,165.95196)
			rgb255=(250.67264999999998,169.274865,168.42418500000002)
			rgb255=(252.81924,171.34444499999998,170.90508)
			rgb255=(254.967615,173.41836,173.38725000000002)
		}
}

\pgfplotsset{
	discard if not/.style 2 args={
			x filter/.code={
					\edef\tempa{\thisrow{#1}}
					\edef\tempb{#2}
					\ifx\tempa\tempb
					\else
						
					\fi
				}
		}
}

\pgfdeclarelayer{background}
\pgfdeclarelayer{foreground}
\pgfsetlayers{background,main,foreground}

\usepgfplotslibrary{external,fillbetween}
\RequirePackage{shellesc}

\usepackage{xcolor}
\definecolor{UKLred} {RGB}{207, 25,  59}
\definecolor{UKLblue}{RGB}{ 47, 63, 157}
\definecolor{NYUpurple} {RGB}{88,15,139}
\definecolor{Pastrami} {RGB}{229,85,79}
\definecolor{TheLake}{RGB}{72,159,223}
\definecolor{EastRiver}{RGB}{0,115,152}
\definecolor{SpicyMustard}{RGB}{203,160,82}
\definecolor{CentralPark}{RGB}{0,108,91}
\definecolor{ProspectPark}{RGB}{64,192,172}
\definecolor{turquois}{rgb}{0,0.75,0.75}%

\definecolor{PTblue}{RGB}{68,119,170}
\definecolor{PTcyan}{RGB}{102,204,238}
\definecolor{PTgreen}{RGB}{34,136,51}
\definecolor{PTyellow}{RGB}{204,187,68}
\definecolor{PTred}{RGB}{238,102,119}
\definecolor{PTpurple}{RGB}{170,51,119}
\definecolor{PTgrey}{RGB}{187,187,187}

\usepackage{colortbl}
\usepackage{upgreek}
\usepackage{xr}
\usepackage{nameref}
\usepackage{currfile}

\makeatletter
\newcommand*{\addFileDependency}[1]{
	\typeout{(#1)}
	\@addtofilelist{#1}
	\IfFileExists{#1}{}{\typeout{No file #1.}}
}
\makeatother

\usepackage{pgfplotstable}

\begin{document}

\title{Sensitivity of literature $T_1$ mapping methods to the underlying magnetization transfer parameters}

\author[1,2]{Jakob Assl\"ander*}{\orcid{0000-0003-2288-038X}}
\authormark{Jakob Assl\"ander}

\address[1]{\orgdiv{Center for Biomedical Imaging, Dept. of Radiology}, \orgname{NYU School of Medicine}, \orgaddress{\state{NY}, \country{USA}}}
\address[2]{\orgdiv{Center for Advanced Imaging Innovation and Research (CAI\textsuperscript{2}R), Dept. of Radiology}, \orgname{NYU School of Medicine}, \orgaddress{\state{NY}, \country{USA}}}

\corres{*Jakob Assl\"ander, Center for Biomedical Imaging, NYU School of Medicine, 227 E 30 Street, New York, NY 10016, USA.\\ \email{jakob.asslaender@nyumc.org}}

\finfo{This work was supported by \fundingAgency{NIH/NINDS} grant \fundingNumber{R01~NS131948} and \fundingAgency{NIH/NIBIB} grant \fundingNumber{P41~EB017183}.}

\abstract[Abstract]{
    \textbf{Purpose:}
    Magnetization transfer (MT) has been identified as the principal source of $T_1$ variability in the MRI literature. This study assesses the sensitivity of established $T_1$ mapping techniques to variations in the underlying MT parameters.

    \textbf{Methods:}
    For each $T_1$-mapping method, the observed $T_1$ was simulated as a function of the underlying MT parameters $p_i^\text{MT}$, corresponding to different brain regions of interest (ROIs) at 3T. As measures of sensitivity, the derivatives $\partial T_1^\text{observed} / \partial p_i^\text{MT}$ were computed and analyzed with a linear mixed-effects model as a function of $p_i^\text{MT}$, ROI, pulse sequence type (e.g., inversion recovery, variable flip angle), and the individual sequences.

    \textbf{Results:}
    The analyzed $T_1$-mapping sequences have a considerable sensitivity to changes in the semi-solid spin pool size $m_0^\text{s}$, $T_1^\text{f}$ of the free, $T_1^\text{s}$ of the semi-solid spin pool, and the (inverse) exchange rate $T_\text{x}$. All derivatives vary considerably with the underlying MT parameters and between pulse sequences. In general, the derivatives cannot be determined by the sequence type, but rather depend on the implementation details of the sequence. One notable exception is that variable-flip-angle methods are, in general, more sensitive to the exchange rate than inversion-recovery methods.

    \textbf{Conclusion:}
    Variations in the observed $T_1$ can be caused by several underlying MT parameters, and the sensitivity to each parameter depends on both the underlying MT parameters and the sequence.
}

\maketitle
\footnotetext{
    \textbf{Abbreviations:}
    MT, magnetization transfer;
    ROI, region of interest;
    MP-RAGE, magnetization-prepared rapid gradient-echo;
    DESPOT, driven equilibrium single pulse observation of T1 and T2;
    WM, white matter;
    GM, gray matter;
    CC, corpus callosum.
    \\ Word Count: 2368}

\section{Introduction}
The workhorses of clinical MRI are $T_1$- and $T_2$-\textit{weighted} sequences that exploit longitudinal and transversal spin relaxation, respectively, to create contrasts between different types of tissue and between healthy tissue and pathology.
To provide reproducible measures, many quantitative MRI approaches aim to quantify $T_1$ and $T_2$, that is, they model the MRI signal with the Bloch equations. \cite{Bloch.1946}
However, a considerable variability of literature $T_1$ values has been reported, ranging from 0.6 to 1.1\,s for brain white matter at 3\,T. \cite{Stikov.2015,Ou.2008,ZavalaBojorquez2016,Teixeira.2019}
In our recent work, a colleague and I attributed the majority of this variability to magnetization transfer, and argued that a mono-exponential $T_1$ in biological tissue should be considered a \textit{semi-quantitative} metric. \cite{Assländer.2025}

Magnetization transfer (MT) describes the interaction between spins associated with liquids and macromolecules. \cite{Wolff1989,Henkelman1993}
MT has traditionally been considered a nuisance effect for $T_1$ mapping, despite the early hypothesis of Koenig et al.~\cite{Koenig1990} that MT is a driver of longitudinal spin relaxation.
More recently, this hypothesis was confirmed by the identification of substantial differences between $T_1$ of the semi-solid (0.34\,s in brain white matter at 3\,T) and the free spin pool (1.84\,s in brain white matter at 3\,T). \cite{Helms2009,Gelderen2016,Assländer.2024elq} From these observations, one can conclude that an MT model describes longitudinal relaxation in tissue more accurately compared to the Bloch model.

Given the large volume of the literature on mono-exponential $T_1$ mapping methods and their undeniable benefits in terms of acquisition speed and simplicity, I here investigate how changes in the MT parameters impact the semi-quantitative metric $T_1^\text{observed}$ ($T_1^\text{o}$ for short) as measured with different pulse sequences.

\section{Methods}
I simulated 25 $T_1$-mapping methods from the literature, grouped into
inversion-recovery, \cite{Stanisz.2005, Stikov.2015, Preibisch.2009ng, Shin.2009, Lu.2005, Reynolds.2023, Wright.2008}
Look-Locker, \cite{Stikov.2015, Shin.2009}
saturation-recovery, \cite{Reynolds.2023}
variable flip angle, \cite{Stikov.2015,Cheng.2006,Preibisch.2009ng,Teixeira.2019,Preibisch.2009,Chavez.2012}
and MP\textsuperscript{(2)}RAGE. \cite{Wright.2008, Marques.2010}
The simulations were performed with the same code used in Ref.~\citen{Assländer.2025}.
To briefly recap, the simulations neglect imaging gradients, assume homogeneous $B_0$ and $B_1^+$ fields, and emphasize an adequate representation of the RF pulses.
Specifically, RF pulses were simulated with the generalized Bloch model \cite{Assländer.2022u6f} to describe the spin dynamics of the semi-solid pool. Shaped RF-pulses were implemented as analytic functions, and the spin dynamics during the RF pulse were calculated by solving the differential equation with the canonical 4\textsuperscript{th}-order Runge-Kutta method, embedded in the method of steps \cite{Bellen.2003} to account for the integro-differential nature of the generalized Bloch model. \cite{Assländer.2022u6f}
Rectangular RF pulses were simulated with matrix exponentiation and a linearization of the generalized Bloch model.
Since the focus of this paper is longitudinal relaxation, I further assume complete spoiling, limiting the analysis of the $T_2^\text{f}$-analysis to effects originating from the MT model.
For each sequence, $T_1^\text{o}$ was estimated from the simulated data with the fitting procedures described in the respective publication.
Each simulation can be summarized as $T_1^\text{o} = f_j(m_0^\text{s}, T_1^\text{f}, T_2^\text{f}, T_\text{x}, T_1^\text{s}, T_2^\text{s})$ where $j$ identifies the pulse sequence and $f_j$ is a function of 6 MT parameters: $m_0^\text{s}$ is the semi-solid spin pool size, $T_{1,2}$ are the relaxation times, qualified by the superscripts $^\text{f,s}$ to identify the free and semi-solid spin pool, respectively, and $T_\text{x}$ is the inverse exchange rate.
Note that the simulation accounts for different $T_1^\text{f}$ and $T_1^\text{s}$ values for the free and the semi-solid spin pools, respectively, in accordance with Refs. \citen{Helms2009,Gelderen2016,Assländer.2024elq}.
The model assumes the normalization $m_0^\text{f} + m_0^\text{s} = 1$, and an overall scaling factor is set to one during the simulation, but fitted along with $T_1^\text{o}$ for each sequence.
Therefore, $T_1^\text{o}$ is not sensitive to the scaling factor in the simulation, and the sensitivity to $m_0^\text{f}$ is represented by the derivative with respect to $m_0^\text{s}$.

The simulations were performed for MT parameters measured at 3\,T (cf. Tab.~2 in Ref.~\citen{Assländer.2024elq}), covering 9 regions of interest (ROIs), namely the entire white matter (WM), anterior corpus callosum (CC), posterior CC, cortical gray matter (GM), caudate, putamen, pallidum, thalamus, and hippocampus.
For each of these ROIs and each of the 25 pulse sequences, I calculated $\partial T_1^\text{o} / \partial p_i^\text{MT}$ for all $p_i^\text{MT} \in \{m_0^\text{s}, T_1^\text{f}, T_2^\text{f}, T_\text{x}, T_1^\text{s}, T_2^\text{s}\}$. Due to the wide variety in sequences and fitting methods, gradients were determined with the 5\textsuperscript{th}-order central finite difference method.

The size of each derivative $\partial T_1^\text{o} / \partial p_i^\text{MT}$ was assessed with its mean absolute value across all ROIs and pulse sequences.
To allow for a comparison between parameters, it is normalized by the mean of the parameter value across all ROIs ($\mu_{|\partial T_1^\text{o} / \partial p_i^\text{MT}|} \cdot \mu_{p_i^\text{MT}}$).
The variability of each derivative was assessed with the coefficient of variation, that is, the standard deviation divided by the mean absolute value, each taken over all ROIs and sequences ($\sigma_{\partial T_1^\text{o} / \partial p_i^\text{MT}} / \mu_{|\partial T_1^\text{o} / \partial p_i^\text{MT}|}$).
Further, I modeled each derivative's variability with a linear mixed effects model, which considers each ROI's set of MT parameters $p_i^\text{MT} \in \{m_0^\text{s}, T_1^\text{f}, T_2^\text{f}, T_\text{x}, T_1^\text{s}, T_2^\text{s}\}$ and an intercept as fixed effects, and the ROI, the sequence type, and the individual sequence as random effects.
The mixed model uses a hierarchical structure for the sequence type, that is, inversion-recovery, Look-Locker, saturation-recovery, variable flip angle, and MP\textsuperscript{(2)}RAGE, and individual sequence identifiers.
The model considers random intercepts, but not random slopes. Further, it assumes Gaussian distributions. I used a maximum likelihood algorithm to fit the model to the data.

The large number of parameters entails the risk of overfitting. Therefore, I considered including only a subset of $p_i^\text{MT} \in \{m_0^\text{s}, T_1^\text{f}, T_2^\text{f}, T_\text{x}, T_1^\text{s}, T_2^\text{s}\}$. Comparing Akaike's Information Criterion between all combinations of parameters revealed that, for most derivatives, the best model includes all parameters.
For a few derivatives, dropping one parameter resulted in a marginally improved information criterion. For simplicity, however, I decided to use the full model for all parameters.

I calculated the coefficient of determination $R^2_\text{full}$ of the full model, that is, considering fixed and random effects, with the method proposed by Nakagawa and Schielzeth. \cite{Nakagawa.2013}
The summands in $R^2_\text{full} = R^2_\text{fixed} + R^2_\text{ROI} + R^2_\text{seq. type} + R^2_\text{ind. seq.}$ were identified by calculating $R^2$ of only fixed effects (all jointly) and each random effect separately with the Nakagawa method.
Here, $R^2_\text{ROI}$ captures the ROI-identifier as a random variable, potentially modeling inter-ROI variations not captured by the fixed effects.
$R^2_\text{seq. type}$ captures the degree to which the sequence type and $R^2_\text{ind. seq.}$ captures each sequence by itself.
$R^2_\text{fixed}$ captures all fixed effects, that is, the degree to which variations of the $p_i^\text{MT}$ between the ROIs explain the derivatives' variability.
I isolated the contributions of each fixed effect, that is, each MT parameter: $R^2_\text{fixed} = R^2_{m_0^\text{s}} + R^2_{T_1^\text{f}} + R^2_{T_2^\text{f}} + R^2_{T_\text{x}} + R^2_{T_1^\text{s}} + R^2_{T_2^\text{s}}$, using Shapley regression. \cite{Budescu.1993, Kruskal.1987, Lipovetsky.2001}
This approach calculates $R^2$ for all combinations of fixed effects to assess each parameter's contribution.
I used Nakagawa's method for this analysis as well, considering the intercept and all random effects in addition to a particular combination of fixed effects.

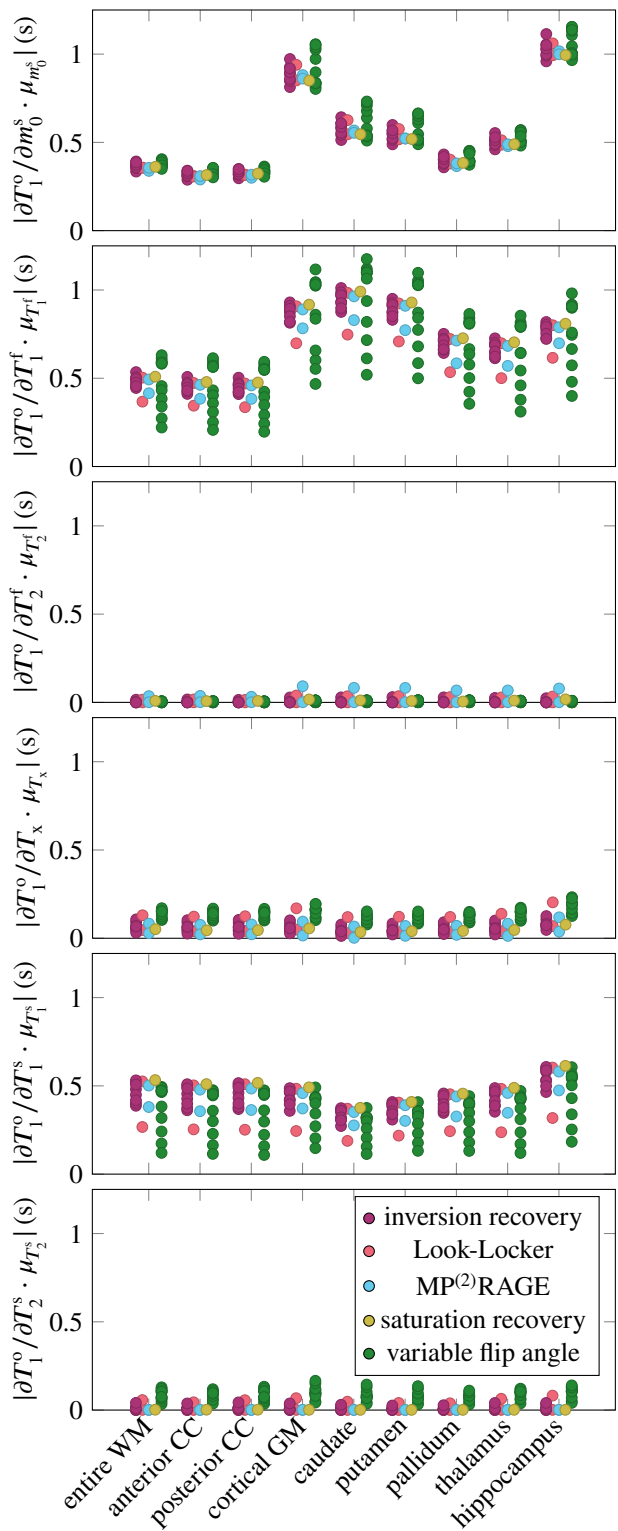
\begin{figure}[htbp]
    \centering
    \begin{tikzpicture}
    \begin{axis}[
            width=0.8\columnwidth,
            height=2.9cm,
            scale only axis,
            ylabel={$|\partial T_1^\text{o} / \partial m_0^\text{s} \cdot \mu_{m_0^\text{s}}|$\,(s)},
            ylabel style = {yshift = -0.2cm},
            xtick={1,...,9},
            xticklabels=\empty,
            ymin = 0,
            ymax = 1.25,
            colormap={bw}{color(0cm)=(PTpurple); color(1cm)=(PTred); color(2cm)=(PTcyan); color(3cm)=(PTyellow); color(4cm)=(PTgreen)},
            name=m0s,
        ]

        \addplot+[
            scatter,
            scatter src=explicit,
            only marks,
            mark=*,
        ] table[
                col sep=comma,
                x=x,
                y=dT1odm0s,
                meta=color,
            ] {\currfiledir/jacobian.csv};

    \end{axis}

    \begin{axis}[
            width=0.8\columnwidth,
            height=2.9cm,
            scale only axis,
            ylabel={$|\partial T_1^\text{o} / \partial T_1^\text{f} \cdot \mu_{T_1^\text{f}}|$\,(s)},
            ylabel style = {yshift = -0.2cm},
            xtick={1,...,9},
            xticklabels= \empty,
            ymin = 0,
            ymax = 1.25,
            colormap={bw}{color(0cm)=(PTpurple); color(1cm)=(PTred); color(2cm)=(PTcyan); color(3cm)=(PTyellow); color(4cm)=(PTgreen)},
            name=T1f,
            at=(m0s.south),
            anchor=north,
            yshift = -0.2cm,
        ]

        \addplot+[
            scatter,
            scatter src=explicit,
            only marks,
            mark=*,
        ] table[
                col sep=comma,
                x=x,
                y=dT1odT1f,
                meta=color,
            ] {\currfiledir/jacobian.csv};

    \end{axis}

    \begin{axis}[
            width=0.8\columnwidth,
            height=2.9cm,
            scale only axis,
            ylabel={$|\partial T_1^\text{o} / \partial T_2^\text{f} \cdot \mu_{T_2^\text{f}}|$\,(s)},
            ylabel style = {yshift = -0.2cm},
            xtick={1,...,9},
            xticklabels= \empty,
            ymin = 0,
            ymax = 1.25,
            colormap={bw}{color(0cm)=(PTpurple); color(1cm)=(PTred); color(2cm)=(PTcyan); color(3cm)=(PTyellow); color(4cm)=(PTgreen)},
            name=T2f,
            at=(T1f.south),
            anchor=north,
            yshift = -0.2cm,
        ]

        \addplot+[
            scatter,
            scatter src=explicit,
            only marks,
            mark=*,
        ] table[
                col sep=comma,
                x=x,
                y=dT1odT2f,
                meta=color,
            ] {\currfiledir/jacobian.csv};

    \end{axis}

    \begin{axis}[
            width=0.8\columnwidth,
            height=2.9cm,
            scale only axis,
            ylabel={$|\partial T_1^\text{o} / \partial T_\text{x} \cdot \mu_{T_\text{x}}|$\,(s)},
            ylabel style = {yshift = -0.2cm},
            xtick={1,...,9},
            xticklabels= \empty,
            ymin = 0,
            ymax = 1.25,
            colormap={bw}{color(0cm)=(PTpurple); color(1cm)=(PTred); color(2cm)=(PTcyan); color(3cm)=(PTyellow); color(4cm)=(PTgreen)},
            name=Tex,
            at=(T2f.south),
            anchor=north,
            yshift = -0.2cm,
        ]

        \addplot+[
            scatter,
            scatter src=explicit,
            only marks,
            mark=*,
        ] table[
                col sep=comma,
                x=x,
                y=dT1odTex,
                meta=color,
            ] {\currfiledir/jacobian.csv};

    \end{axis}

    \begin{axis}[
            width=0.8\columnwidth,
            height=2.9cm,
            scale only axis,
            ylabel={$|\partial T_1^\text{o} / \partial T_1^\text{s} \cdot \mu_{T_1^\text{s}}|$\,(s)},
            ylabel style = {yshift = -0.2cm},
            xtick={1,...,9},
            xticklabels= \empty,
            ymin = 0,
            ymax = 1.25,
            colormap={bw}{color(0cm)=(PTpurple); color(1cm)=(PTred); color(2cm)=(PTcyan); color(3cm)=(PTyellow); color(4cm)=(PTgreen)},
            name=T1s,
            at=(Tex.south),
            anchor=north,
            yshift = -0.2cm,
        ]

        \addplot+[
            scatter,
            scatter src=explicit,
            only marks,
            mark=*,
        ] table[
                col sep=comma,
                x=x,
                y=dT1odT1s,
                meta=color,
            ] {\currfiledir/jacobian.csv};

    \end{axis}

    \begin{axis}[
            width=0.8\columnwidth,
            height=2.9cm,
            scale only axis,
            ylabel={$|\partial T_1^\text{o} / \partial T_2^\text{s} \cdot \mu_{T_2^\text{s}}|$\,(s)},
            ylabel style = {yshift = -0.2cm},
            xtick={1,...,9},
            xticklabels={entire WM, anterior CC, posterior CC, cortical GM, caudate, putamen, pallidum, thalamus, hippocampus},
            xticklabel style = {anchor = north east, rotate=45, xshift = 0cm, yshift = 0.15cm},
            ymin = 0,
            ymax = 1.25,
            colormap={bw}{color(0cm)=(PTpurple); color(1cm)=(PTred); color(2cm)=(PTcyan); color(3cm)=(PTyellow); color(4cm)=(PTgreen)},
            name=T2s,
            at=(T1s.south),
            anchor=north,
            yshift = -0.2cm,
            legend,
            legend pos = north east,
            legend style = {
                    fill opacity = 0.6,
                    draw opacity = 1,
                    text opacity = 1,
                },
        ]

        \addlegendimage{legend image code/.code={\draw[fill=PTpurple, draw=black, fill opacity = 1] (0cm,0cm) circle (0.07cm);}}
        \addlegendentry{inversion recovery}
        \addlegendimage{legend image code/.code={\draw[fill=PTred, draw=black, fill opacity = 1] (0cm,0cm) circle (0.07cm);}}
        \addlegendentry{Look-Locker}
        \addlegendimage{legend image code/.code={\draw[fill=PTcyan, draw=black, fill opacity = 1] (0cm,0cm) circle (0.07cm);}}
        \addlegendentry{MP\textsuperscript{(2)}RAGE}
        \addlegendimage{legend image code/.code={\draw[fill=PTyellow, draw=black, fill opacity = 1] (0cm,0cm) circle (0.07cm);}}
        \addlegendentry{saturation recovery}
        \addlegendimage{legend image code/.code={\draw[fill=PTgreen, draw=black, fill opacity = 1] (0cm,0cm) circle (0.07cm);}}
        \addlegendentry{variable flip angle}

        \addplot+[
            scatter,
            scatter src=explicit,
            only marks,
            mark=*,
        ] table[
                col sep=comma,
                x=x,
                y=dT1odT2s,
                meta=color,
            ] {\currfiledir/jacobian.csv};

    \end{axis}
\end{tikzpicture}
    \vspace{-0.25cm}
    \caption{
        Absolute value of the observed $T_1^\text{o}$'s derivative with respect to the 6 MT parameters.
        I calculated the derivatives for 9 brain regions of interest (ROIs) and for 25 pulse sequences, grouped into different sequence types (cf. legend).
        Here, WM denotes white matter, CC the corpus callosum, and GM gray matter.
        The derivatives $\partial T_1^\text{o} / \partial p_i^\text{MT}$ are normalized by $p_i^\text{MT}$, averaged over all ROIs, to allow for a comparison between the parameters.
    }
    \label{fig:derivatives_ROI}
\end{figure}

\section{Results}
\begin{figure}[b!]
    \centering
    \input{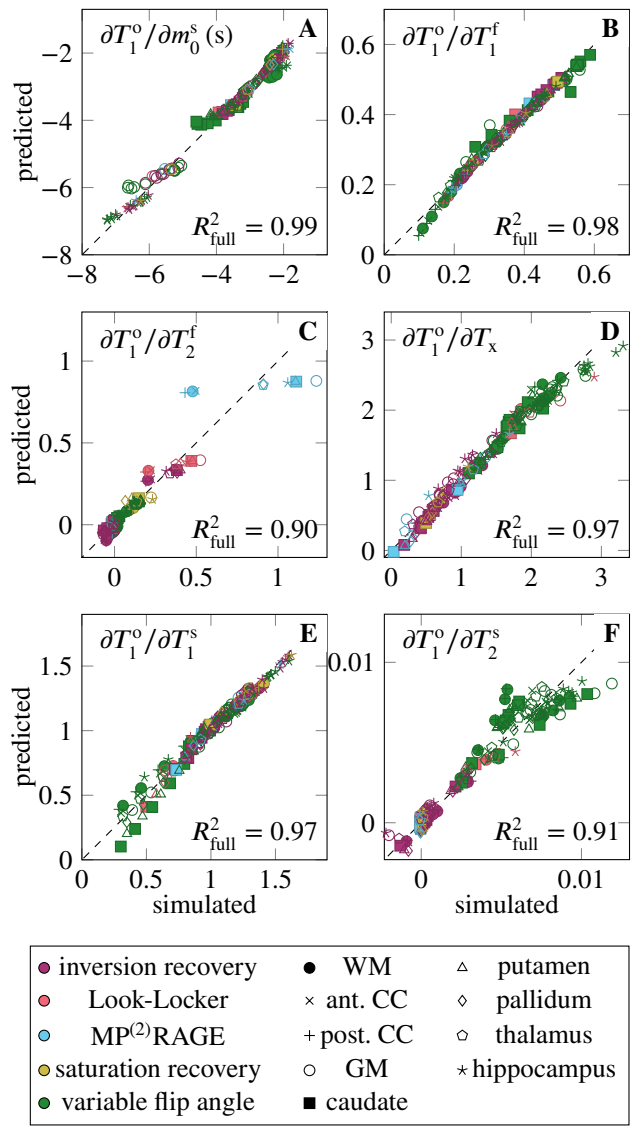}
    \vspace{-0.25cm}
    \caption{Validation of the mixed effects model, where ``simulated'' denotes the simulated derivatives, and ``predicted'' denotes the output of the mixed model.
        The sequence type is here color-coded, while the maker shape identifies the region of interest.
        Here, WM denotes white matter, CC the corpus callosum, and GM gray matter.
        The dotted line represents the perfect fit.
        For convenience, the $R^2_\text{full}$ is denoted in the bottom right of each plot, repeating the values shown in Tab.~\ref{tab:R2_full}.
        For an interactive version of this plot, where the sequence type and ROI can be toggled for display, refer to the link in the data availability statement.
    }
    \label{fig:prediction}
\end{figure}

Fig.~\ref{fig:derivatives_ROI} aggregates the six derivatives $\partial T_1^\text{o} / \partial p_i^\text{MT}$ for all ROIs and pulse sequences. On average, the pulse sequences are most sensitive to $T_1^\text{f}$, closely followed by $m_0^\text{s}$ and $T_1^\text{s}$.
The sequences also have a notable sensitivity to $T_\text{x}$, while they are less sensitive to $T_2^\text{f}$ and $T_2^\text{s}$ (cf. Tab.~\ref{tab:R2_full}, column $\mu_{|\partial T_1^\text{o} / \partial p_i^\text{MT}|} \cdot \mu_{p_i^\text{MT}}$).
The mixed model explains the variability of all derivatives well, with $R^2_\text{full} \in [0.90, 0.99]$ (Fig.~\ref{fig:prediction}).
None of the derivatives are sensitive to the ROI as a random effect, suggesting that the linear model of the MT parameters (fixed effects) sufficiently describes the differences between the ROIs.
The following paragraphs discuss the derivatives in the order listed in Tab.~\ref{tab:R2_full}.

The derivative $\partial T_1^\text{o} / \partial m_0^\text{s}$ is large on average, and shows considerable variability, which is quantified by the coefficient of variation $\sigma_{\partial T_1^\text{o} / \partial m_0^\text{s}} / \mu_{|\partial T_1^\text{o} / \partial m_0^\text{s}|} = 0.44$ (cf. Tab.~\ref{tab:R2_full}).
The mixed model regression reveals that the vast majority of the variability is explained by the fixed effects, that is, by the MT parameters (column $R^2_\text{fixed}$ in Tab.~\ref{tab:R2_full}).
Examining the fixed effects further shows that $m_0^\text{s}$, $T_1^\text{f}$, and $T_2^\text{f}$ jointly explain approximately 80\% of $R^2_\text{fixed}$ (Tab.~\ref{tab:fixed_effects}).
Despite the excellent $R^2_\text{full} = 0.99$, there is structure in the residuals (Fig.~\ref{fig:prediction}A). Specifically, the slope of the predicted vs. simulated derivative is lower in the GM, hippocampus, and caudate, which might be overcome by a non-linear model.

The derivative $\partial T_1^\text{o} / \partial T_1^\text{f}$ is, on average, the largest derivative, but only by a 21\% margin compared to $\partial T_1^\text{o} / \partial m_0^\text{s}$.
Its variability ($\sigma_{\partial T_1^\text{o} / \partial T_1^\text{f}} / \mu_{|\partial T_1^\text{o} / \partial T_1^\text{f}|} = 0.33$) is, to approximately 2/3, explained by the MT parameters (fixed effects), and to 1/3 by individual sequence identifier (Tab.~\ref{tab:R2_full}).
Among the fixed effects, $m_0^\text{s}$ and $T_1^\text{s}$ play the biggest role (Tab.~\ref{tab:fixed_effects}).

The derivative $\partial T_1^\text{o} / \partial T_2^\text{f}$ is small, except for the MP\textsuperscript{2}\-RAGE \cite{Marques.2010} (Fig.~\ref{fig:derivatives_ROI}).
Note, however, that this analysis focuses on the MT model and neglects coherence pathways (it assumes perfect spoiling), which contribute considerably to the sensitivity of $T_1$ mapping methods to $T_2^\text{f}$, in particular for variable-flip-angle methods. \cite{Preibisch.2009ng, Yarnykh.2010, Stikov.2015, Heule.2016, Baudrexel.2018, Corbin.2021}
Since the MT model is ill-equipped to analyze $\partial T_1^\text{o} / \partial T_2^\text{f}$, I will not discuss it further and refer to the literature.

\begin{table*}[t!]
    \newcommand{\g}{\cellcolor[gray]{0.8}}
    \begin{center}
        \begin{tabular}[]{c|c|c|c|c|c|c|c}
            $\partial T_1^\text{o} / \partial p_i^\text{MT}$ & $\mu_{\left| \frac{\partial T_1^\text{o}}{\partial p_i^\text{MT}} \right|} \cdot \mu_{p_i^\text{MT}}$ & $\sigma_{\frac{\partial T_1^\text{o}}{\partial p_i^\text{MT}}} / \mu_{\left| \frac{\partial T_1^\text{o}}{\partial p_i^\text{MT}} \right|}$ & $R^2_\text{fixed}$ & $R^2_\text{ROI}$ & $R^2_\text{seq. type}$ & $R^2_\text{ind. seq.}$ & $R^2_\text{full}$ \\
            \midrule
            $\partial T_1^\text{o} / \partial m_0^\text{s}$  & 0.56                                                                                                  & 0.44                                                                                                                                        & 0.96               & 0.00             & 0.00                   & 0.02                   & 0.99              \\
            $\partial T_1^\text{o} / \partial T_1^\text{f}$  & 0.68                                                                                                  & 0.33                                                                                                                                        & 0.66               & 0.00             & 0.00                   & 0.32                   & 0.98              \\
            $\partial T_1^\text{o} / \partial T_2^\text{f}$  & 0.01                                                                                                  & 1.96                                                                                                                                        & 0.02               & 0.00             & 0.20                   & 0.68                   & 0.90              \\
            $\partial T_1^\text{o} / \partial T_\text{x}$    & 0.10                                                                                                  & 0.51                                                                                                                                        & 0.11               & 0.00             & 0.49                   & 0.36                   & 0.97              \\
            $\partial T_1^\text{o} / \partial T_1^\text{s}$  & 0.39                                                                                                  & 0.28                                                                                                                                        & 0.23               & 0.00             & 0.01                   & 0.72                   & 0.97              \\
            $\partial T_1^\text{o} / \partial T_2^\text{s}$  & 0.05                                                                                                  & 0.98                                                                                                                                        & 0.01               & 0.00             & 0.63                   & 0.26                   & 0.91              \\
            \bottomrule
        \end{tabular}
    \end{center}
    \caption{Mixed effects model analysis of the derivatives.
        The column $\mu_{|\partial T_1^\text{o} / \partial p_i^\text{MT}|} \cdot \mu_{p_i^\text{MT}}$ denotes the mean derivative, normalized by the average parameter, and serves as a measure for the sensitivity of $T_1^\text{o}$ to the respective parameter.
        The column $\sigma_{\partial T_1^\text{o} / \partial p_i^\text{MT}} / \mu_{|\partial T_1^\text{o} / \partial p_i^\text{MT}|}$ denotes the coefficient of variation.
        The coefficients of determination for the full model $R^2_\text{full}$ is dissected into its components: $R^2_\text{full} = R^2_\text{fixed} + R^2_\text{ROI} + R^2_\text{seq. type} + R^2_\text{ind. seq.}$, where
        $R^2_\text{fixed}$ captures all fixed effects, that is, the degree to which variations of the $p_i^\text{MT}$ between the ROIs explain the derivatives' variability.
        $R^2_\text{ROI}$ captures the ROI-identifier as a random variable, potentially modeling inter-ROI variations not captured by the linear model of $p_i^\text{MT}$.
        $R^2_\text{seq. type}$ captures the degree to which the sequence type, that is, the groups inversion-recovery, Look-Locker, saturation-recovery, variable flip angle, and MP\textsuperscript{(2)}RAGE, explains variability of the derivatives, and $R^2_\text{ind. seq.}$ captures each sequence by itself.
    }
    \label{tab:R2_full}
\end{table*}

The derivative $\partial T_1^\text{o} / \partial T_\text{x}$ is (normalized) the fourth largest. Its considerable variability ($\sigma_{\partial T_1^\text{o} / \partial T_\text{x}} / \mu_{|\partial T_1^\text{o} / \partial T_\text{x}|} = 0.51$) is, unlike the other derivatives (except for $\partial T_1^\text{o} / \partial T_2^\text{s}$), foremost explained by the sequence type, followed by the individual sequence, while the fixed effects play a subordinate role.
Fig.~\ref{fig:prediction}D gives some insights into the structure of the sequence-type effect. Variable-flip-angle methods appear to be rather sensitive to $T_\text{x}$, while the examined inversion-recovery, saturation-recovery, and MP\textsuperscript{(2)}RAGE methods are less sensitive. The two Look-Locker sequences have rather different behavior, where the version proposed by Stikov et al. \cite{Stikov.2015} has high sensitivity and the version proposed by Shin et al. \cite{Shin.2009} has a low sensitivity.

\begin{table}[t!]
    \newcommand{\g}{\cellcolor[gray]{0.8}}
    \setlength{\tabcolsep}{5pt}
    \begin{center}
        \begin{tabular}[]{c|c|c|c|c|c|c}
            $\partial T_1^\text{o} / \partial p_i^\text{MT}$ & $R^2_{m_0^\text{s}}$ & $R^2_{T_1^\text{f}}$ & $R^2_{T_2^\text{f}}$ & $R^2_{T_\text{x}}$ & $R^2_{T_1^\text{s}}$ & $R^2_{T_2^\text{s}}$ \\
            \midrule
            $\partial T_1^\text{o} / \partial m_0^\text{s}$  & 0.24                 & 0.33                 & 0.20                 & 0.07               & 0.10                 & 0.03                 \\
            $\partial T_1^\text{o} / \partial T_1^\text{f}$  & 0.24                 & 0.06                 & 0.03                 & 0.05               & 0.18                 & 0.10                 \\
            $\partial T_1^\text{o} / \partial T_2^\text{f}$  & 0.01                 & 0.00                 & 0.00                 & 0.00               & 0.00                 & 0.00                 \\
            $\partial T_1^\text{o} / \partial T_\text{x}$    & 0.01                 & 0.04                 & 0.03                 & 0.00               & 0.01                 & 0.02                 \\
            $\partial T_1^\text{o} / \partial T_1^\text{s}$  & 0.02                 & 0.06                 & 0.05                 & 0.01               & 0.04                 & 0.04                 \\
            $\partial T_1^\text{o} / \partial T_2^\text{s}$  & 0.00                 & 0.01                 & 0.00                 & 0.00               & 0.00                 & 0.00                 \\
            \bottomrule
        \end{tabular}
    \end{center}
    \caption{Analysis of the fixed effects. $R^2_\text{fixed}$ is separated into the individual effects with Shapley regression.\cite{Budescu.1993, Kruskal.1987, Lipovetsky.2001} Note that $R^2_\text{fixed} = R^2_{m_0^\text{s}} + R^2_{T_1^\text{f}} + R^2_{T_2^\text{f}} + R^2_{T_\text{x}} + R^2_{T_1^\text{s}} + R^2_{T_2^\text{s}}$.
    }
    \label{tab:fixed_effects}
\end{table}

The derivative $\partial T_1^\text{o} / \partial T_1^\text{s}$ is the third largest and has a moderate variability ($\sigma_{\partial T_1^\text{o} / \partial T_1^\text{s}} / \mu_{|\partial T_1^\text{o} / \partial T_1^\text{s}|} = 0.28$). The variability is foremost explained by the individual sequences, followed by the MT parameters (in particular the four relaxation times $T_{1,2}^{\text{f,s}}$).
Analyzing the residuals in Fig.~\ref{fig:prediction}E, we find that most deviations from the mixed effects model stem from variable-flip-angle methods and Stikov's Look-Locker sequence.
The largest deviations occur at small derivatives and are dominated by the putamen and caudate, which are underestimated, and by the WM, corpus callosum, and hippocampus, which are overestimated.

The derivative $\partial T_1^\text{o} / \partial T_2^\text{s}$ is the second smallest derivative (7\% of $\partial T_1^\text{o} / \partial T_1^\text{f}$ when normalized).
Its large relative variation ($\sigma_{\partial T_1^\text{o} / \partial T_2^\text{s}} / \mu_{|\partial T_1^\text{o} / \partial T_2^\text{s}|} = 0.98$) is almost fully explained by the pulse sequence, with the type explaining about 2/3. Fig.~\ref{fig:prediction}F shows that variable-flip-angle methods are most sensitive to $T_2^\text{s}$, but there is a large variability within this group of sequences. Inversion recovery methods show a smaller sensitivity, also with considerable variability that spreads beyond the origin.
The examined MP\textsuperscript{(2)}RAGE, saturation recovery, and Shin's Look-Locker implementation have the lowest sensitivity, while Stikov's Look-Locker implementation has a higher sensitivity, in the range of variable-flip-angle methods.

\section{Discussion}
The purpose of this paper is to study the relation between the MT and the mono-exponential models and to relate changes in the observed $T_1^\text{o}$ to changes in the underlying MT parameters. We find that $T_1^\text{o}$ is sensitive to all parameters that model the dynamics of the coupled spin-system's longitudinal magnetization ($m_0^\text{s}$, $T_1^\text{f}$, $T_\text{x}$, and $T_1^\text{s}$).
In contrast, $T_1^\text{o}$ shows low sensitivity to $T_2^\text{f,s}$.

An analysis of the four most relevant MT parameters revealed a considerable variability with coefficients of variation $\in [0.28, 0.51]$. This implies that the composition of sensitivities varies between sequences and tissues, hampering broad-stroke conclusions.

The derivative with respect to $m_0^\text{s}$ depends foremost on the MT parameters themselves; that is, one could use Tab.~\ref{tab:fixed_effects_model} to predict the derivative, assuming the MT parameters are known.
With reduced accuracy (cf. the ratio $R^2_\text{fixed}/R^2_\text{full}$ in Tab.~\ref{tab:R2_full}), the same approach could be used for $T_1^\text{f}$. Note that $T_1^\text{f}$ and $m_0^\text{s}$ are, on average, the two largest derivatives.

In contrast, the derivative with respect to $T_\text{x}$ can be predicted based on the sequence, largely independent of the MT parameters. The sequence type provides partial information---variable-flip-angle methods are more sensitive than others---but a better prediction can be achieved when sequence details are known.
The sensitivity to $T_1^\text{s}$ has a slightly higher, but still low, dependence on the MT parameters. The sequence type is, however, not informative, and only an analysis of a specific implementation allows for a reasonable prediction of the sensitivity.

\subsection{Study limitations}
My goal was to include sufficient sequences to assess the impact of the sequence type.
This study includes 12 implementations of the variable-flip-angle and 8 of the inversion-recovery method, which allows for careful conclusions about typical implementations of these types.
However, an individual implementation might still be an outlier.
For example, I determined that inversion-recovery methods are, in general, less sensitive to $T_\text{x}$, but this sensitivity increases when measuring short inversion times. \cite{Gochberg.2003}
In contrast, the simulation includes only 1 saturation-recovery, 2 Look-Locker, and 2 MP\textsuperscript{(2)}RAGE sequences, which do not allow for drawing conclusions beyond the particular implementations.
This is exemplified by the, in parts, diverging sensitivities of the two Look-Locker methods.

This paper focuses on the most widespread $T_1$ mapping methods.
In multi-parameter methods, the estimate of one parameter depends, in general, on the other parameters.
For example, in DESPOT, \cite{Deoni.2003} the estimation of $T_2$ is explicitly a function of $T_1$.
In MR Fingerprinting, \cite{Ma2013} this dependency is implicit.
Therefore, it is expected that $T_2$-estimates with multi-parametric methods are also sensitive to MT parameters, which is confirmed by Ref. \citen{Hilbert.2019}.
A more detailed analysis of multi-parameter methods is, however, beyond the scope of this paper.

The linear fit of the fixed parameters explains the variability, arguably, surprisingly well. However, based on the known underlying physics, we expect poor performance when extrapolating outside the examined parameter space.
For example, at $m_0^\text{s} = 0$, the MT model reduces to the Bloch model, and we would expect $\partial T_1^\text{o} / \partial T_1^\text{f} = 1$ and  $\partial T_1^\text{o} / \partial p_i^\text{MT} = 0$ for $p_i^\text{MT} \in \{ T_\text{x}, T_1^\text{s}, T_2^\text{s}\}$.
The rightmost column in Tab.~\ref{tab:fixed_effects_model} demonstrates that this is not the case, highlighting the limitations of this linear approximation.

As mentioned before, the simulations do not include coherence pathways, which, in reality, contribute substantially to the derivative $\partial T_1^\text{o} / \partial T_2^\text{f}$. I refer, e.g., to Refs.~\citen{Preibisch.2009ng, Yarnykh.2010, Stikov.2015, Heule.2016, Baudrexel.2018, Corbin.2021} for detailed discussions of this topic.

\subsection{Outlook}
The purpose of this paper is to provide an overview of $T_1^\text{o}$'s sensitivities to changes in the underlying MT parameters. Pathologies or healthy tissue changes commonly entail changes in multiple MT parameters. The sensitivity of a $T_1$ mapping sequence to such tissue changes can be assessed by summing the derivatives, weighted by the expected parameter changes. A more detailed analysis of specific changes can be performed by tailoring the provided source code accordingly.

This paper simulates pulse sequences and protocols described in the literature and analyzes $T_1^\text{o}$'s sensitivities to changes in the underlying MT parameters. A complementary analysis would be to analyze the sensitivity of $T_1^\text{o}$ to individual parameters in the protocol of a particular pulse sequence. While such an analysis is beyond the scope of this paper, the provided source code can easily be modified to perform such an analysis.

\section{Conclusion}
The analyses in this paper confirm the general assumption that, in most cases, $T_1^\text{observed}$ is foremost sensitive to $T_1^\text{f}$ of the free spin pool.
However, the derivatives with respect to $m_0^s$ and $T_1^\text{s}$ are only marginally smaller.
All derivatives vary considerably as a function of the MT parameters and the pulse sequence.
The MT parameters themselves matter in particular for $\partial T_1^\text{o} / \partial m_0^\text{s}$ and $\partial T_1^\text{o} / \partial T_1^\text{f}$, which can be predicted with the linear regression model (Tab.~\ref{tab:fixed_effects_model}).
The sequence matters in particular for $\partial T_1^\text{o} / \partial T_\text{x}$ and $\partial T_1^\text{o} / \partial T_1^\text{s}$ and, in general, the sequence type is not enough to determine the sensitivities, and implementation details need to be considered.
The only exception is that variable-flip-angle sequences are, in general, more sensitive to exchange than inversion-recovery sequences.

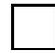
\appendix
\renewcommand{\thetable}{A\arabic{table}}

\begin{table*}[t!]
    \newcommand{\g}{\cellcolor[gray]{0.8}}
    \begin{center}
        \begin{tabular}[]{c|c|c|c|c|c|c|c|c}
            $\partial T_1^\text{o} / \partial p_i^\text{MT}$ & $a_0$ & $a_{m_0^\text{s}}$ & $a_{T_1^\text{f}}$\,(1/s) & $a_{T_2^\text{f}}$\,(1/s) & $a_{T_\text{x}}$\,(1/s) & $a_{T_1^\text{s}}$\,(1/s) & $a_{T_2^\text{s}}$\,(1/$\upmu$s) & $m_0^\text{s} = 0$ \\
            \midrule
            $\partial T_1^\text{o} / \partial m_0^\text{s}$  & 3.27  & 5.54               & -2.98                     & -39.15                    & 61.87                   & -0.69                     & -0.21                            & -4.38              \\
            $\partial T_1^\text{o} / \partial T_1^\text{f}$  & 0.22  & -1.44              & -0.16                     & 3.62                      & -2.02                   & 0.83                      & 0.02                             & 0.57               \\
            $\partial T_1^\text{o} / \partial T_2^\text{f}$  & 0.17  & -0.38              & 0.03                      & -1.24                     & -0.13                   & 0.09                      & 0.00                             & 0.19               \\
            $\partial T_1^\text{o} / \partial T_\text{x}$    & 1.11  & 1.41               & 0.91                      & -3.20                     & -5.17                   & -2.99                     & -0.02                            & 0.93               \\
            $\partial T_1^\text{o} / \partial T_1^\text{s}$  & 1.52  & 1.10               & 0.55                      & -2.40                     & -4.95                   & -2.56                     & -0.02                            & 0.86               \\
            $\partial T_1^\text{o} / \partial T_2^\text{s}$  & -0.00 & 0.00               & 0.00                      & -0.02                     & 0.02                    & 0.00                      & -0.00                            & 0.00               \\
            \bottomrule
        \end{tabular}
    \end{center}
    \caption{Coefficients $a_{p_i^\text{MT}}$ of the fixed effects for each derivative's mixed model fit, including the intercept $a_0$.
        The last column denotes the derivatives assuming $m_0^\text{s} = 0$ and the mean values for the other parameters.
        For $m_0^\text{s} = 0$, the MT model reduces to the Bloch model, and a perfect statistical model should result in $\partial T_1^\text{o} / \partial T_1^\text{f} = 1$ and $\partial T_1^\text{o} / \partial T_\text{x} = \partial T_1^\text{o} / \partial T_1^\text{s} = \partial T_1^\text{o} / \partial T_2^\text{s} = 0$. This is clearly not the case, highlighting the limitations of the mixed-effects model when extrapolating.
    }
    \label{tab:fixed_effects_model}
\end{table*}

\section{Acknowledgements}
The author would like to thank Drs.
Stanisz,
Stikov, Boudreau, Leppert,
Marques,
Malik, Teixeira,
Michal, Reynolds,
Van Zijl,
Cheng,
Gowland,
Preibisch, Deichmann,
and Shin
for providing unpublished implementation details of their $T_1$-mapping methods.


\section{Data availability statement} \label{sec:data_availability}
Code to replicate all results can be found at \url{https://github.com/JakobAsslaender/T1variability}.
The results in the present paper were created with v2.0 of the simulation code.
The website \url{https://jakobasslaender.github.io/T1variability/v2.0/} documents the code and outlines all simulation code along with the presented results.
It also replicates the figures of this paper with interactive features, such as hovers that allow for the identification of the pulse sequence that corresponds to each data point.
Clicking on legend entries also allows for the analysis of individual sequence groups.


\bibliography{My_Library.bib}
\end{document}